\documentclass[a4paper,10pt]{article}
\usepackage{t1enc}

\begin{document}

\begin{center}
\noindent
ON THE QUESTION OF VALIDITY \\ OF THE ANTHROPIC PRINCIPLES

\vspace*{4mm}
\noindent
Zsolt Hetesi, B\'ela Bal\'azs \\ E\"otv\"os University Dept. of Astronomy \\
H-1117, Budapest P\'azm\'any walk 1/a \\
zs.hetes@astro.elte.hu, bab@ludens.elte.hu
(Acta Physica Polonica B Vol. \textbf {37.} No. 9. 2729-2739)
\end{center}

\vspace*{4mm}
\noindent
{\small {\textbf {Abstract}}

\noindent
During the last centuries of human history, many questions was repeated in connection with the great problems of the existence and origin of human beings, and also of the Universe. The old questions of common sense and philosophy have not been solved in spite of the indisputable results of modern natural sciences. Recently the so-called anthropic principles show that these questions are still present.

We investigated some important results of the modern cosmology and their consequences with respect to the corresponding questions of philosophy and logic. After a short conceptual introduction there are two baselines. It is showed first how Goedel's theroem affects the foundation of anthropic principles. Our train of thought shows that Goedel's incompleteness theorem may deny some efforts claiming that anthropic principles can be ruled out. After this in the Appendix we touch the branch of questions that are connected with the philosophical aspects of anthropic principles and the multiple-world hypothesis. Here we investigated those formulae of quantum theory, which are supposed to be the ground for the theory of many worlds-hypothesis.

Although our method is based partly on philosophy and logic, it is mainly grounded in the results and methods of the natural sciences. So we need both physics and philosophy to go in our way.

\noindent
\textbf{Keywords}: anthropic principles, physics and philosophy, mathematical logic.}

\section{Introduction}
In this paper after a short historical and conceptual introduction (Sec. 2) we will investigate our main thesis in Sec. 3.: our paper is connected in general with the origin of anthropic principles which can be interpreted in a philosophical manner: which incentives produced them, when and in what circumstances will their effect be questionable, or when does this effect disappear? Our purpose is to find an answer to these questions. However the main goal is the next. We will show that the validity of the anthropic principles cannot be denied making use of the unified theories (GUT, ToE). To prove our thesis we will also use Goedel's theorem. In Appendices 1. and 2. we will examine in a comprehensive sense the physical background and in some aspects the philosophical background of the anthropic principles, and their consequences in philosophical fields - and naturally we will give the correspondent definitions of them.

\section{Concepts}

\subsection{Introduction of the Anthropic Principles on the basis of their historical background}
As science has grown, the elements of the dominant medieval world-concepts have disappeared, one after the other, resulting in the following conclusions:
\vspace*{4mm}
\begin{itemize}
\item{the Earth is not in the center of the world, and it has not a significant place in the Universe,}
\item{the heavens are not built of concentric spheres, and the motion of celestial bodies is not regular (i.e. it can not be described by spherical motions),}
\item{the structure of the Universe is developed through physical processes,}
\item{living beings are a result of an evolution.\footnote{We do not accept the equivalence of different world-concepts or various systems of cosmological explanations, and we think that more modern physical theories could better approach the objective reality of the world.}}
\end{itemize}

\vspace*{4mm}
Up to the present, science, and several branches of philosophy of science, have tried to eliminate the Creator on the basis of scientific theories. It seemed that a mechanistic world-concept which dominated the physics of the 19th century could be a complete explanation of the Universe. Many scientists have expected the end of physical research.\footnote{They thought that there were only some unsolved problems which could soon be solved. One of these problems   was the blackbody radiaton, from which the quantum-theory has emerged. As it is known the quantum-theory has not yet reached its final state. Recently there have been similar opinions in connection with the ''final theory'', (i.e. that the end of physics is imminent).} These expectations were first confuted by the uncertainty principle of quantum-mechanics.\footnote{According to the uncertainty principle  in physics it is not possible to measure with infinite precision the value of two so-called conjugate quantities. Thus the uncertainty  in the measurement of the position of an electron varies inversely with  the uncertainty in the measurement of its momentum.} Afterwards Goedel's theorem of incompleteness verified that the axiomatization of mathematics was useless, i.e. the axiomatic mathematics cannot completely explain all the phenomena of the Universe. Finally, in informatics Turing's theorem restricted the possibilities of the automata.

This development reached its culmination in 1973, when Brandon Carter defined his famous principle, which was intended as a methodological cosmological principle. He called it the anthropic principle, and he distinguished between two versions of this principle: the weak and the strong anthropic principle (abbreviated WAP and SAP).  Its weak form is presented by Carter in the following manner:

\begin{quote}
All the cosmological theories have to take into account the fact that our location in the Universe is necessarily privileged to the extent of being compatible with our existence as observers. (Carter, 1974)
\end{quote}

This statement contains the essential elements. However, we have to give a fuller definition of the WAP as follows:

\begin{quote}
The observed values of all physical and cosmological quantities are not equally probable, but they take on values restricted by the requirement that sites can be found where carbon based life can evolve, and by the requirement  that the Universe is old enough for it to have already done so (Barrow-Tipler 1996, p. 16.).
\end{quote}

The WAP states that the physical constants and laws of the Universe must be such as we measure them, i.e. these laws and constants must be compatible with our life. The WAP was used in this meaning during its history. For example in the 60's when Hoyle, Bondi and Gold developed the steady-state cosmological model stating the continuous creation of matter and giving an alternative to the standard cosmological model. Before it was disproved by observational data in its original form, the use of WAP had shown the improbability of the steady-state cosmological theory.\footnote{As it was shown by M. Rees (1974), in the standard cosmology the timescale of stellar evolution is in the order of Hubble time. There is quite little chance in a steady-state cosmology to be the same.}
Brandon Carter stated also the strong anthropic principle (SAP). We will use the definition of the SAP as follows:

\begin{quote}
The Universe must have those properties which allow life to develop within it at some stage in its history.\footnote{Interesting to note that according to a more fine division there is an ordering and an allowing strong anthropic principle (SAP1 and SAP2). If the first is true, the life is necessarily emerging in a fine-tuned universe, but in the allowing one only the possibility emerges. In logical view the true SAP is the SAP1, and SAP2 is closer to the WAP. According to Carter's original statement the strong version is the following: ''The Universe (and hence the fundamental parameters on which it depends) must be such as to admit the creation of intelligent observers within it at some stage of its evolution.'' (Carter 1974, p. 294.)} (Barrow-Tipler 1996, p. 21.)
\end{quote}

We note that this is more speculative than WAP. In the SAP Carter argued the ''explanation'' of why the values of dimensionless physical constants are what they are. The immediate consequence of SAP is that the physical laws and constants must be such as to allow the emergence of life. For further definitions see the Appendices and Bal\'azs, (2005), Hetesi-V\'egh (2006) .

\subsection{Unified theories in physics}

When physics successfully constructed a complete theory, scientists immediately started to investigate if a new theory was suitable for the line of already existing models and processes or if a contradiction emerged. If a new theory passes the test,
it must be the result of valid physical thinking. In former times the situation was that in some areas of phenomena the theories which explained them, brought to a higher level, could also serve as a universal model (unified theory) explaining all phenomena.

An obvious example was the unification of electricity and magnetism done by Maxwell.  As it is known, Maxwell's four equations describe all electromagnetic processes, and if conditions are satisfactory they split into two pairs of equations: electrostatic and magnetic equations.

      Up to the present physics has found four fundamental forces: gravitational force, electromagnetic force, weak and strong interactions. The main purpose of recent investigations is to unify these forces in a grand unification. The unification of electro-magnetic and weak forces was a success (with electroweak force as a result), and later succeeded to incorporate the strong interaction, but it had not yet contained the gravitational force. The theory which unifies the three mentioned forces is called Grand Unified Theory (GUT).

After the success of GUT a group of theoretical physicists thought that it would be possible to construct such a theory which will be able to describe all physical phenomena and which shall contain all the former theories. It is called ''world-formula'' or Theory of Everything (ToE).

\section{Possible consequences of a ''world-formula''. \\ Goedel's theorem}

As we have mentioned in Part 2, some physicists, among those who are working on the grand unification, hope that the ''world-formula'' can be found. It will be able to explain everything, it will also serve as explanation of the existence of the Universe, and it will show that the world is necessarily this and cannot be other. The latter statement is true because a final world-formula is deductive (one say), every other thing is only a corollary of it.
After such expectations some scientists declared that unified theories will put an end to the questions which lead us from the fact of the fine-tuned Universe to the anthropic principles (Kane et al. 2002). If these expectations could be fulfilled, then it would be a complete world-explanation, and it would signify the end of the anthropic principles, because a ToE puts an end to the fact of the fine-tuned physical constants and laws.

However, these hopes are excessive for two reasons:

\begin{itemize}
\item{Goedel's theorem denies that an arithmetical system which is at least as large as the system of {\it Principia Mathemtica} is complete in itself.}
\item{There is no physical theory which can explain why the world exists just so.
Goedel (1931) proved his undecideness or incompleteness theory which (with some simplification) states the following two theorems:}
\end{itemize}

\begin{quote}
First theorem: If an axiomatic theory of sets is contradiction-free, then there are sentences which are neither provable nor unprovable.

Second theorem: In any consistent axiomatizable theory the consistence of the system is not provable in the system.{\footnote {See Myers (2000) for details.}}
\end{quote}

From these two theorems we can draw the following important conclusions. There is no mathematical theorem which carries its own trueness in itself while the set of axioms, which are the base of this theorem, is uncertain with regard to its self-consistence. But theoretical physics uses mathematics to describe the world. From this follow the facts mentioned below.
      Because of Goedels theorem, a reliable final world-formula cannot be constructed. If it were realizable then it would be true, but not yet necessarily as it was observed by Jaki (1987), (1998, pp. 89-117). (Note that Hawking, a former adherent of a ToE, also discovered Goedel's theory and its relevance in physics. (Hawking, 2000) (Hawking, 2003)).
      However Jaki's train of thought needs some explanation.  Let us start with our key statement. If an
axionmatized mathematical theory which is equivalent to ToE is
contradiction-free then it is not complete, and if it is complete then it
cannot be contradiction-free.

    Also if ToE can be made then it must be axiomatizable because otherwise
it is not assured that it would include all physical laws. As Tipler states (Tipler 1990, p.323.) any scientific theory -- which is a logical system indeed -- is based on axioms. Axioms can be modified, also the andvance is always possible for the present axioms always may be found to be consequences of more fondamental axioms. If this is the case theories may more and more convergent. 

It seems in the history of physics that the more forces and phenomena is included the less freedom remains to construct the new unified theory. If we suppose that this advance reaches its end i.e. ToE is constructed, it is no doubt, that ToE is also axiomatic. Tipler notes that Goedel's theorem acts on ToE iff ToE is at least difficult as the system of {\it Principia Mathetica}. According to him it is "quite possible that a ToE could lie in one of the decidable branches of mathematics." (Tipler, 1990, p. 325.) \footnote{There are several branches of mathematics which can be proven decidable and consistent by reference to itself.}

We think that Tipler's argument is incorrect. If a ToE is ToE it contains all the mathematics hence it will be at least as difficult as the number theory (i.e. the {\it Principia Mathematica}). Furthermore the physical world includes at least natural numbers, and it is described by a system of words which can be translated into a formal physical theory. Because number theory is based on the natural numbers, phyisical theories are at least as large as {\it Principia Mathematica} (Kitada, 1999).

Also if ToE is
axiomatized, it will be at least as complex as the system of  {\it Principia Mathematica}, so Goedel's theorem can be applied to it:  If the axiomatized
system which is equivalent to ToE is contradiction-free, then it is not
complete, hence there will be statements within the system that are neither
provable nor unprovable. Thus it cannot represent a theory of everything
which is contrary to our supposition.\footnote{In the case of a ToE completeness and contradiction-freeness is a necessary condition!} In connection with this thought there is a remarkable sentence in a paper of Bal\'azs: ''In this case physics would not be an empiric science at all but a part of the deductive logic. Everything can be deduced from unambiguous axioms which were verified by the theory itself and which could not be traced back to other known natural laws. As Steven Weinberg has written, if the Theory of Everything anyway could  be realizable then it would have had to be logically isolated, i.e. it would not be modifiable because it would collapse then.'' (Bal\'azs 2001) In any case, by Goedels theorem we come to the remarkable conclusion that a possible ToE does not preclude the anthropic principles.

Our second argument is the following. Let us suppose (even if impossible) that someone will construct a world-formula and nothing is in conflict with it, i.e. every phenomenon fits in it. As Stephen Hawking rightly has noted, world-formula is nothing but just masses of equations and laws. Anthropic principles are usable in the future too because every number and formula is specific enough for it to be obvious from one glimpse, in other words: a hypothetic final answer in physics, like a ''world-formula'', only strengthens the contingency of the world because it leads to the question: why is the world as we find it and not something different.

\section{Conclusion}
As we have seen anthropic principles emerged because state that physical constants and laws of the Universe must be compatible with our life. Hence these principles contain philosophical contents (see Appendix 2.). But in special cases these contents can be different for different scientists. We showed that Goedel's incompleteness theroem may deny those efforts which aim to show that anthropic principles can be ruled out by some form of unified theories. Interpretation of Goedel's theorem in physics have started just now (except Tipler's short reflection on Jaki's article (Jaki, 1987), Tipler, 1990 p. 325) hence uses of Goedel's theorem in physics are obvious today but its exact place is not. Or in other words: different approaches can be equally valid.

{\small
\subsection*{Appendix 1. Further anthropic definitions}
In the main text we mentioned the definitions of WAP and SAP only. Here we detail these definitions and also investigate further definitions.
There exist several different versions or supplements of SAP. The most known is the design argument (DA):

\begin{quote}
There exits one possible universe ''designed'' with the goal of generating and sustaining ''observers''. (Barrow-Tipler 1996, p. 22.)
\end{quote}

This DA seems to be neither probable nor deniable in any physical or logical way, and it has a strong religious content when the ''design'' is a work of a Designer. Hence physicists do not use it willingly (however there exist some exceptions). Instead of the DA, several variants have appeared. All of them try to answer question of the connection between human existence and fine-tuned physical constants detailed in SAP to avoid in any way the design argument (DA).

The participatory anthropic principle (PAP) appears as a result of the Copenhagen interpretation of quantum-mechanics:

\begin{quote}
Observers are necessary to bring the universe into being. (Barrow-Tipler 1996, p. 22.)
\end{quote}

Now let us consider a train of thought which drives us to the third variant of SAP called many worlds-hypothesis (MWH). To state that the physical constants of the universe derive specific values from an ensemble of different values, we have to suppose the (i) conceptual or (ii) real existence of the numerical ensemble.

The DA has chosen the first solution (i), when the other possibilities exist only as possibilities in the mind of the Designer.

The many-words-hypothesis represents the second version (ii), the strict definition of which is the following:

\begin{quote}
An ensemble of other different universes is necessary for the existence of our Universe. (Barrow-Tipler 1996, p. 22.)
\end{quote}

If we think PAP to be true, then we will have the following question: what happens when life (emerged according to SAP) once dies out? Does the universe itself cease to exist, because of a lack of quantum-effect of observers?
The final anthropic principle (FAP) was born to solve this problem:

\begin{quote}
In the Universe intelligent information-processing life must come into existence, and once it comes into existence it will never die out. (Barrow-Tipler 1996, p. 23.)
\end{quote}

It is worth emphasizing that only the WAP seems to be well-founded among the above definitions.

\subsection*{Appendix 2.Philosophical reflections on hidden philosophical contents in anthropic principles}
As it was partly stated in Appendix 1., except for WAP's methodological character, these principles have speculative and philosophical contents. In connection with the design argument (DA) we have mentioned that it does not seem to be provable or deniable by mathematical or scientific methods, hence physicists usually do not accept it. Now let us examine some philosophical contents of anthropic principles in detail.\footnote{Being strict, these statements often are only non-physical. But we do not claim non-physical and philosophical statements to be equivalent.}

\vspace*{4mm}
\noindent
SAP contains a powerful but physically not provable statement. The use of the anthropic argument in SAP has ontological meaning. In our point of view, physics is not able to say anything about existence, only about quantitative aspects of existing things.

\vspace*{4mm}
\noindent
DA has ontological meaning too, insofar as it refers to a Designer. It is usually refuted because of its philosophical content. In opinions about SAP this problem does not rise.

\vspace*{4mm}
\noindent
PAP is based on the Copenhagen interpretation of quantum-mechanics. This interpretation is accepted by the majority of physicists but only in methodological sense, a minority considers it in ontological meaning. In its methodological meaning the Copenhagen interpretation states only that e.g. there is no separate particle or wave which can be measured, just wave-pockets. This treatment identifies physical reality with the mathematical model describing it. However, it does not consider this identity true in an ontological meaning, but as a working hypothesis. The trouble starts when this methodological conception is extended to the existence of things. It is true e.g. that two quantities having a certain connection between them are not measurable ''with punctuality''at the same time, but it is not valid to state that these quantities do not exist ''punctually'' just because they are not measurable ''punctually''.\footnote{In quantum mechanics measurement means that the measured system has more eigenvalues and it jumps in one of them during the measurement. Systems exist in the linear combination of the all possible state and not in eigenstate before measuring. Mathematically: $\psi=\sum_{i=1}^n|c_i|^2 a_i$, where $\psi$ is the wave function of the system, $a_i$ is the value of the i-th eigenstate, $|c_i|^2$ is the probability of i-th state. All $|c_i|^2$ are different and non-zero before measurement, but only one of them will be non-zero and has the value of 1 after the measurement. $\sum|c_i|^2=1$ is always true, i.e. the system is complete. Interesting to note that the Copenhagen interpretation has a rival theory, the theory of hidden variables. According to von Neumann's proof (von Neumann 1955 [1932]) there are not dispersion-free states thus a theory of hidden variables is impossible. Bell showed that von Neumann's proof is false (Bell 2004 [1964]). There is an interesting corollary in the theory of hidden variables. In order to accept a theory of hidden variables one has to accept the violation of locality. Latest experiments have confirmed that the Copenhagen interpretation of quantum-mechanics is not satisfactory (Home-Gribbin 1999).} This latter statement is the fallacy of petitio principii.\footnote{There is a supposition in the background (only measurable things can exist) which is not a proven fact. The petitio principii fault is in the inadequate use of the world ''punctually''. When it refers to the measurement it has operational meaning and when it refers to the existence it has ontological meaning.}(Jaki, 1998, pp. 117-147.)

Those who, according to the PAP, argue that universe needs observers because observers contract the wave-functions of it and with this brings it into existence, the Copenhagen interpretation is used in its second (and defective) meaning. There is no doubt that it is possible to construct a continuity equation from wave-function, and the obtained $|\psi|^2$ streams in time, and it is also true that $|\psi|^2$  more and more flows away in time, but it is wrong to conclude from this that an away-flowing wave-pocket can be contracted to a particle (or to Universe in this case)  for two reasons:

\begin{itemize}
\item{The meaning of the square of wave-function is the probability-density of finding the particle. It is not identical with the particle. The particle does not flow away, but our knowledge about the locality of the particle becomes more and more uncertain in time. (The probability-density is a mathematical construction and is not able to contract.)\footnote{These statements are only true if we do not choose Copenhagen interpretation as philosophical background.}}
\item{To contract the Universe's wave-function one must observe the Universe from an external point, but it is per definitionem impossible in the case of the Universe.}
\end{itemize}

\vspace*{4mm}
\noindent
Concerning MWH some of the problems are caused by the very idea of the ensemble. Hence questions which emerge here are not purely ontological but also logical and methodological. If there are many universes (with a small u), then they either interact with each other or not. If yes (they interact), then there is only one Universe, if not, then it is meaningless to speak about them from a physical point of view because they are not observable.\footnote{If there is no interaction MWH does not has other sign which is observable in our Universe. The existence of the interaction does not seem to be possible because the MWH hypothesis states that physical constants and/or laws are different in different universes.} However, their existence is not inconsistent with this chain of ideas, because it would be a logical fault as it was mentioned in the criticism of PAP. The many-words-hypothesis is valid for a physicist only if there can be interaction between the various universes of the MWH, even if, yet as of today this is not observable. Furthermore it is questionable whether there can be any meaningful use of statistics where there is only one element (namely our own world) observable in that hypothetical ensemble.

Finally the origin of the existence of this ensemble is also questionable. If physical constants and laws are different in every universe, then what laws create the ensemble itself? Can these laws be called physical laws?

\vspace*{4mm}
\noindent
Concerning the FAP: it has a strong ontological character, and it cannot be verified physically or mathematically.}

\vspace*{4mm}
\noindent
{\small {\textbf {Acknowledgement:}} The authors thank Gustav Teres SJ, L\'aszl\'o V\'egh and Ervin Nemesszeghy SJ
for useful and friendly discussions and proposals. Technical help of
Andr\'as P\'al and Emese Forg\'acs-Dajka is also kindly appreciated.}

\vspace*{4mm}
\noindent
{\small Bibliography

\vspace*{4mm}
\noindent
Bal\'azs, B (2005). The cosmological replication cycle, the extraterrestrial paradigm and the final anthropic principle. Diotima, 2005 Vol. 33. pp. 44-453

\vspace*{1mm}
\noindent
Bal\'azs, B (2001). Anthropic principles and the family tree of the universes.\\ http://astro.elte.hu/~bab/seti/IACP12z.htm

\vspace*{1mm}
\noindent
Barrow, J.-Tipler, F.(1996). The anthropic cosmological principle. Oxford: Oxford Univ. Press, 2nd ed.

\vspace*{1mm}
\noindent
Bell, J. S.(2004). On the problems of hidden variables in quantum mechanics. \\ In Speakable and Unspeakable in Quantum Mechanics pp. 1-13. Cambridge: Cambridge University Press,  3rd ed. (first published 1964)

\vspace*{1mm}
\noindent
B. Carter (1974). Large number coincidences and the anthropic principle in Cosmology. In IAU Symposium no. 63, \\ Confrontation of cosmological theories with observational data. p.293. Ed. M.S.Longair. Reidel, Dordrecht

\vspace*{1mm}
\noindent
Hawking, S. W. (2000). Goedel and the end of physics. Lecture on Dirac's Centenary. \\ http://www.damtp.cam.ac.uk/strtst/dirac/hawking/

\vspace*{1mm}
\noindent
Hawking, S. W. (2003). Goedel and the end of physics. Lecture at Texas A\&M's College of Sciences. \\ http://www.physics.sfasu.edu/astro/news/20030308news/StephenHawking20030308.htm

\vspace*{1mm}
\noindent
Hetesi, Zs.-V\'egh, L. (2006). A defintion for fine tuning. in prep.

\vspace*{1mm}
\noindent
Home, D.-Gribbin, J. (1991). What is light? New Scientist 02 November 1991, Magazine issue 1793

\vspace*{1mm}
\noindent
Jaki, S. L. (1987). Teaching transcendence in physics. Am. J. Phys. {\textbf 55} (10) October 1987 884-888

\vspace*{1mm}
\noindent
Jaki, S. L. (1998). God and the Cosmologists. Michigan: Real View Books, 2nd ed.

\vspace*{1mm}
\noindent
Kane, G. L.; Perry, M. J.; Zytkow, A. N. (2002). The beginning of the end of the anthropic principle. New Astronomy Vol. 7. Issue 1. January 2002, pp. 45-53

\vspace*{1mm}
\noindent
Kitada, H.(1999). A possible solution for the non-existence of time. arXiv:gr-qc/9910080 v3

\vspace*{1mm}
\noindent
Myers, D.(2000). Goedel's Incompleteness Theorem \\
http://www.math.hawaii.edu/~dale/godel/godel.html

\vspace*{1mm}
\noindent
von Neumann, J. (1955). Matematische Grundlagen der Quanten-mechanik  (English transl.: Princeton, N. J.: Princeton University Press, 1955) (first published Berlin: Springer-Verlag, 1932)

\vspace*{1mm}
\noindent
Rees, M. J. (1972). Comm. Astrophys. Space Phys. 4, 182

\vspace*{1mm}
\noindent
Tipler, F. (1990). Modell of an evolving God. In Physics Philosophy and Theology: A Common Quest for Understanding pp. 313- 331 Eds: Russel- Stroeger-Coyne. Vatican City, Vatican Observatory}

\end{document}